\documentclass[pra,aps,twocolumn,nopacs,superscriptaddress,nofootinbib,reprint]{revtex4}
\usepackage{graphicx}  
\usepackage{dcolumn}  
\usepackage{bm}           
\usepackage{amsmath}
\usepackage{epsfig}
\usepackage{indentfirst}
\usepackage{psfrag}
\usepackage{subfigure}
\usepackage{amssymb}
\usepackage{tabularx,multirow}
\usepackage{color}
\usepackage[colorlinks,linkcolor=blue,citecolor=blue,urlcolor=blue,hyperindex,driverfallback=dvipdfm]{hyperref}
\usepackage[T1]{fontenc}
\def\ii{{\rm i}}  

\def\Te{{T_{\rm e}}}   
          
\begin{document}
\title{Ultrafast Topological Engineering in Metamaterials}
\author{Renwen~Yu}
\email{renwen.yu@icloud.com}
\affiliation{ICFO-Institut de Ciencies Fotoniques, The Barcelona Institute of Science and Technology, 08860 Castelldefels (Barcelona), Spain}
\author{Rasoul~Alaee}
\affiliation{Department of Physics, University of Ottawa, Ottawa, ON K1N 6N5, Canada}
\author{Robert~W.~Boyd}
\affiliation{Department of Physics, University of Ottawa, Ottawa, ON K1N 6N5, Canada}
\affiliation{The Institute of Optics, University of Rochester, Rochester, New York 14627, USA}
\author{F.~Javier~Garc\'{\i}a~de~Abajo}
\email{javier.garciadeabajo@nanophotonics.es}
\affiliation{ICFO-Institut de Ciencies Fotoniques, The Barcelona Institute of Science and Technology, 08860 Castelldefels (Barcelona), Spain}
\affiliation{ICREA-Instituci\'o Catalana de Recerca i Estudis Avan\c{c}ats, Passeig Llu\'{\i}s Companys 23, 08010 Barcelona, Spain}
 
\date{\today}
\begin{abstract}
Transient optical heating provides an efficient way to trigger phase transitions in naturally occurring media through ultrashort laser pulse irradiation. A similar approach could be used to induce topological phase transitions in the photonic response of suitably engineered artificial structures known as metamaterials. Here, we predict a topological transition in the isofrequency dispersion contours of a layered graphene metamaterial under optical pumping. We show that the contour topology transforms from elliptic to hyperbolic within a subpicosecond timescale by exploiting the extraordinary photothermal properties of graphene. This new phenomenon allows us to theoretically demonstrate applications in engineering the decay rate of proximal optical emitters, ultrafast beam steering, and dynamical far-field subwavelength imaging. Our study opens a disruptive approach toward ultrafast control of light emission, beam steering, and optical image processing.
\end{abstract}
\maketitle

\section*{Introduction}

Transient heating induced by laser pulse absorption has been intensely studied to induce phase transitions of interest such as metal-insulator in VO$_2$ \cite{XBV11}, amorphous-crystalline in Ge$_2$Sb$_2$Te$_5$ \cite{FWN00}, and charge density waves in 1{\it T}-TaS$_2$ \cite{HEE16}, all of which hold strong potential for applications in next-generation electronic and optical data storage \cite{WY07}. An interesting possibility arises when considering phase transitions in components of artificial structures known as metamaterials, which grant us access into a broader range of optical properties extending beyond those encountered in naturally occurring materials. Actually, fascinating applications have been demonstrated by using metamaterial properties, including negative refraction \cite{V1968,P00}, superlensing \cite{FLS05}, optical cloaking \cite{SMJ06}, and superfocusing \cite{LBP02}. More recently, hyperbolic metamaterials have been found to exhibit an effective uniaxial anisotropic permittivity tensor that produces peculiar hyperbolic topology in the isofrequency dispersion contours and enables engineering of the decay rates of optical emitters \cite{KJN12,LKF14}, as well as hyperbolic waveguiding \cite{PN05} and far-field subwavelength imaging \cite{JAN06,LLX07}, among other feats. Hyperbolic metamaterials can be realized, for example, by simply stacking layers composed by alternating dielectric and plasmonic materials \cite{LKF14,KJN12,PIB13}. Among the latter, graphene offers additional appealing properties compared with noble metals, such as remarkably low optical losses \cite{WLG15,NMS18}, exceptionally small electronic heat capacity \cite{ESG18}, extraordinary photothermal response \cite{paper313}, and the ability to tune its optical conductivity through electrical gating \cite{NGM04,FV07,FP07_2,CGP09}. In fact, hyperbolic metamaterials based on layered graphene/dielectric stacks \cite{IMS13,OGC13} are being intensely explored as an excellent platform for applications involving infrared light \cite{BPA19}.

When subject to ultrafast optical pulse irradiation, the optical energy absorbed by graphene is first deposited in its conduction electrons, which can be heated to an elevated temperature that remains during $\sim0.5-1\,$ps before transferring a substantial fraction of heat to the lattice \cite{JUC13,GPM13,RWW10}. Notably, due to the small heat capacity of electrons in graphene \cite{paper286,ESG18} compared with the lattice, the latter remains close to ambient temperature for optical pump-pulse fluences as high as $\sim2\,$mJ/cm$^2$ and peak electron temperatures of 1000's\,K \cite{LMS10,paper330}, partially assisted by the weak electron-phonon coupling that characterizes this material \cite{ESG18,paper313}. A strong thermo-optical response is then triggered thanks to the strong temperature dependence of the graphene optical conductivity. Consequently, we expect that topological transitions of isofrequency contours could be triggered in metamaterials formed by layered graphene/dielectric stacks under ultrafast optical pumping, similar to light-driven phase-transitions in natural materials.  

\begin{figure*}
\noindent \begin{centering}
\includegraphics[width=0.85\textwidth]{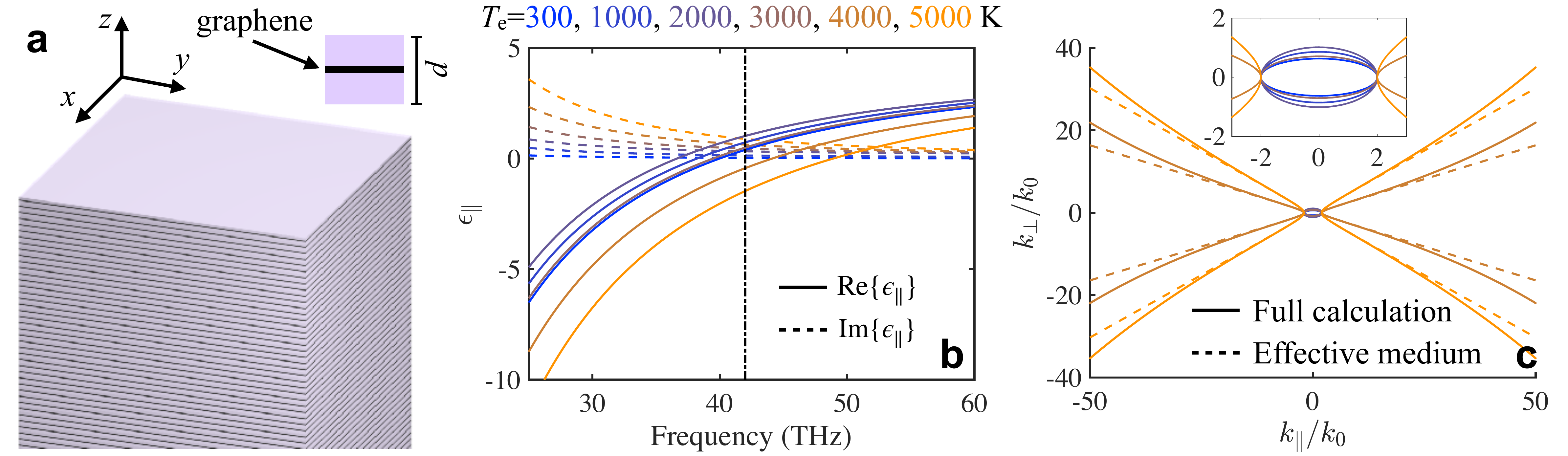}
\par\end{centering}
\caption{(a) Schematic of a metamaterial made of graphene/dielectric stacks. The right inset shows a vertical unit cell with graphene located in the center. In this work, we take a unit cell thickness $d=20\,$nm, a permittivity of the dielectric $\epsilon_{{\rm d}}=4$, and a graphene Fermi energy $E_{{\rm F}}=0.4\,$eV. (b) Real and imaginary parts of the effective in-plane (i.e., for polarization in the $x$-$y$ plane) permittivity $\epsilon_{\parallel}$ as a function of optical frequency at different graphene electron temperatures $T_{{\rm e}}$, as indicated by labels. (c) Isofrequency dispersion contours at a fixed frequency of 42\,THz [indicated as a dash-dotted vertical line in panel (b)] for different electron temperatures. We compare results obtained from electromagnetic numerical simulations (solid curves) and an effective medium model (dashed curves). The topology of the contour transforms from elliptic to hyperbolic as $T_{{\rm e}}$ increases.}
\label{Fig1}
\end{figure*}

In this Letter, we theoretically investigate ultrafast photothermal manipulation of a topological transition in layered metamaterials composed by graphene/dielectric stacks, driven by light absorption in graphene. Based on a realistic description of the temperature-dependent optical properties of graphene, in combination with the spatiotemporal heat flow within its electron/lattice subsystems under ultrafast laser pumping, we predict a topological transition of the isofrequency dispersion contours of the metamaterial in the infrared domain, whereby the topology transforms from elliptic to hyperbolic within an ultrafast timescale. We show that the spatiotemporal dynamics of this topological transition can be probed by a delayed free electron beam, and further find the transient hyperbolic phase to last $\sim1\,$ps. Our results enable several exotic phenomena in the ultrafast regime, such as dynamical engineering of the decay rate of optical emitters in the vicinity of the metamaterial, as well as directional beam steering by carving the metamaterial into a cylindrical lens, which we show to be useful for subwavelength far-field image encoding/decoding.

\section*{Results}

In Fig.\ \ref{Fig1}(a), we sketch a metamaterial composed of graphene/dielectric stacks with graphene located in the center of the dielectric-graphene-dielectric unit cell (see inset). We take a unit cell thickness $d=20\,$nm, a permittivity of the dielectric material $\epsilon_{{\rm d}}=4$, and a graphene Fermi energy $E_{{\rm F}}=0.4\,$eV throughout this work. Given the small value of $d$ compared with the infrared light wavelengths considered in this work, an effective permittivity should accurately describe the dielectric response of the metamaterial. Because of the system symmetry, we then have an effective uniaxial anisotropic material with two different permittivities $\epsilon_{\parallel}$ and $\epsilon_{\perp}$ along in-plane ($x$-$y$ plane) and out-of-plane ($z$ axis) directions, respectively. Additionally, due to the two-dimensional (2D) nature of graphene, we have $\epsilon_{\perp}=\epsilon_{{\rm d}}$. It is well known that optical pulse pumping can elevate the temperature of graphene electrons up to $\sim5000\,$K within a ultrafast timescale \cite{JUC13,GPM13,TSJ13} without causing any damage to the material because of its exceptionally small electronic heat capacity \cite{ESG18}. We present the spectral dependence of the in-plane effective permittivity $\epsilon_{\parallel}=\epsilon_{{\rm d}}+\ii 4\pi\sigma/\omega d$ for different electron temperatures $\Te$ in Fig.\ \ref{Fig1}(b), where $\sigma$ is the temperature-dependent surface conductivity of graphene and $\omega$ is the angular frequency. It should be noted that the epsilon-near-zero frequency (${\rm Re}\{ \epsilon_{\parallel} \}=0$, see solid curves) is first redshifted and then blueshifted as $\Te$ is increased, which is a consequence of the nontrivial $\Te$-dependence of the graphene chemical potential \cite{paper313}. More specifically, the ${\rm Re}\{ \epsilon_{\parallel} \}=0$ condition leads to a frequency $\approx\sqrt{4e^2\mu^{\rm D}/\hbar^2\epsilon_{{\rm d}}d}$, where $\mu^{\rm D}$ is the temperature-dependent effective Drude weight in the graphene conductivity \cite{paper313}. Additionally, ${\rm Im}\{ \epsilon_{\parallel} \}$ (dashed curves) increases with $\Te$ as a consequence of the enhancement in the inelastic scattering rate of graphene electrons \cite{YYM19}. 

We now explore the isofrequency dispersion contours at different electron temperatures for p-polarized electromagnetic fields, which is shown in Fig.\ \ref{Fig1}(c) at a frequency of 42\,THz [indicated by a vertical black dash-dotted line in Fig.\ \ref{Fig1}(b)]. The figure reveals two distinct regimes represented by the topology of the isofrequency contour, which evolves from elliptic to hyperbolic as $\Te$ increases. Specifically, within the $\Te=300-3000\,$K range, the isofrequency contour remains elliptic, despite some variations in shape. When $\Te$ is further increased, a dramatic variation of the contour topology occurs, which results in an emerging hyperbolic metamaterial because ${\rm Re}\{ \epsilon_{\parallel} \}<0$ when $\Te>3000\,$K. These results demonstrate that controlling $\Te$ in graphene can be an efficient and ultrafast route toward inducing topological transitions through optical pulse pumping. Incidentally, we compare the isofrequency dispersion contour calculated with the transfer matrix method in Fig.\ \ref{Fig1}(c) (solid curves) with that obtained from the effective medium theory (dashed curves), as determined by
\begin{align}
\frac{k_{\parallel}^2}{\epsilon_{\perp}}+\frac{k_{\perp}^2}{\epsilon_{\parallel}}=k_0^2,
\label{disp}
\end{align}
where $k_0$ is the free-spade light wave vector, while $k_{\parallel}$ and $k_{\perp}$ are the wave vectors along in- and out-of plane directions, respectively. We attribute the discrepancies between these two methods observed at large values of $k_{\parallel}$ to the lack of validity of the effective medium model when $k_{\parallel}d \ll 1$ no longer holds.

\begin{figure*}
\noindent \begin{centering}
\includegraphics[width=0.85\textwidth]{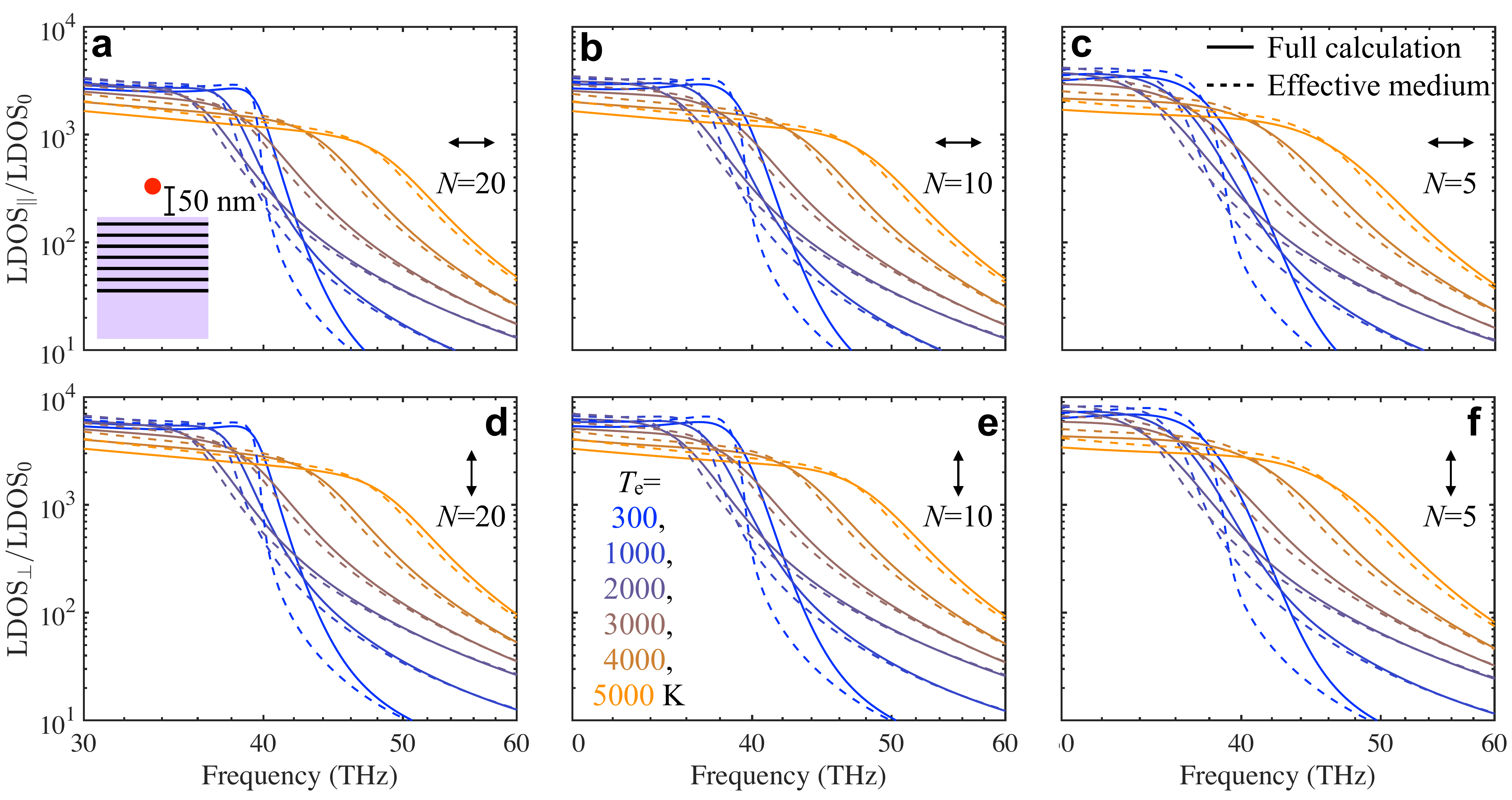}
\par\end{centering}
\caption{(a-c) Temperature-dependent enhancement of the local density of optical states (LDOS) as a function of frequency for metamaterials composed of different numbers $N$ of vertical periods. The probing point dipole has in-plane polarization and is placed 50\,nm away from the surface. (d-f) Same as (a-c) for out-of-plane polarization. We compare the results obtained from a fully numerical calculation (solid curves) with the effective medium model (dashed curves).}
\label{Fig2}
\end{figure*}

A promising application of hyperbolic metamaterials relates to their ability to manipulate the decay rate of proximal emitters by enhancing the local density of optical states (LDOS) \cite{LKF14}, which is given by \cite{LK1977,NH06}
\begin{align}
\frac{{\rm LDOS}_{\parallel}}{{\rm LDOS}_0}=1+\frac{3}{4}\int_{0}^{\infty}\frac{k_xdk_x}{k_0^3}{\rm Re} \left \{\left (\frac{k_0^2 r_s}{k_z}-k_z r_p \right ) {\rm e}^{2\ii k_z z_0}\right \}
\label{LDOS1}
\end{align}
and
\begin{align}
\frac{{\rm LDOS}_{\perp}}{{\rm LDOS}_0}=1+\frac{3}{2}\int_{0}^{\infty}\frac{k_x^3dk_x}{k_0^3}{\rm Re} \left \{ \frac{r_p}{k_z} {\rm e}^{2\ii k_z z_0} \right \}
\label{LDOS2}
\end{align}
for in- and out-of-plane polarization, respectively, where $k_z=\sqrt{k_0^2-k_x^2}$, $z_0$ is the separation distance between the emitter and the metamaterial surface, and $r_s$ and $r_p$ are the reflection coefficients at the metamaterial-air interface for s- and p-polarization. These expressions are normalized to the projected LDOS in free space ${\rm LDOS}_0=\omega^2/3\pi^2 c^3$, where $c$ is the speed of light. The temperature-dependence of the LDOS enhancement is presented in Fig.\ \ref{Fig2} at a distance $z_0=50\,$nm above the upper surface of the metamaterial [see inset in Fig.\ \ref{Fig2}(a)]. A large LDOS enhancement ($>10^3$) is found as a result of strong confinement of the photonic modes supported by the metamaterial. In general, as the electron temperature $\Te$ increases, the spectral range with high LDOS extends to lower frequencies. More specifically, an enhancement of two orders of magnitude can be obtained at $\sim 48\,$THz when increasing $\Te$ (see color-coded labels) due to the topological transformation of the isofrequency contour described in Fig.\ \ref{Fig1}. This means that one can control the emitter decay rate in an ultrafast manner through rising the electron temperature by means of optical pulse pumping. Our findings are robust against the number $N$ of vertical unit cells composing the metamaterial film, as shown in Fig.\ \ref{Fig2}(a-c) for a dipolar emitter polarized parallel to the surface, as calculated from Eq.\ \ref{LDOS1}. Similar results and conclusions are found for out-of-plane polarization, as shown in Fig.\ \ref{Fig2}(d-f), calculated from Eq.\ \ref{LDOS2}. The deviation between the results calculated by using the transfer matrix method (solid curves) or the effective medium model (dashed curves) are again related to the mismatch at large values of $k_{\parallel}$.     

\begin{figure}
\noindent \begin{centering}
\includegraphics[width=0.5\textwidth]{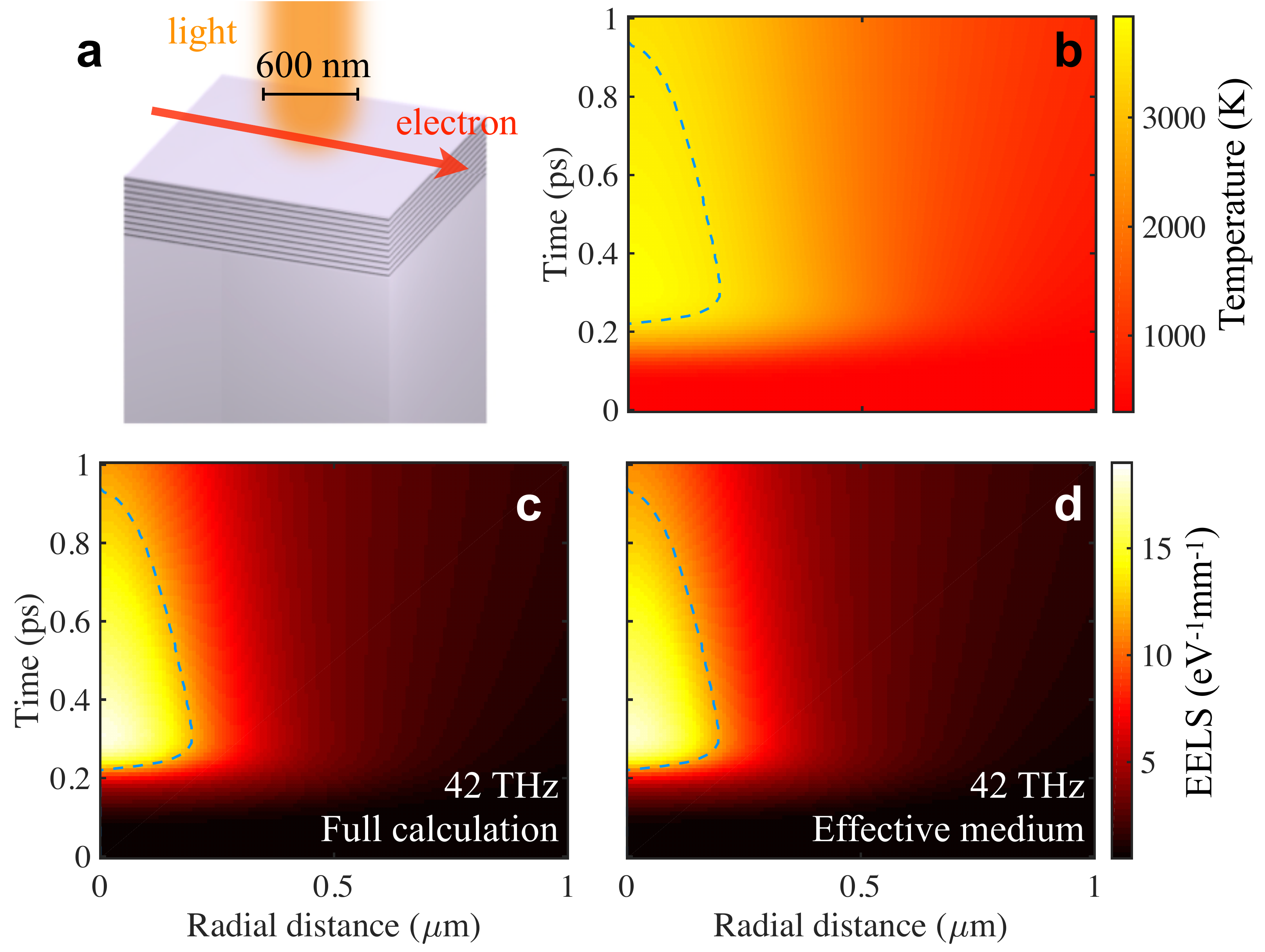}
\par\end{centering}
\caption{(a) Schematic of a pump-probe configuration involving a normally-incident optical pump pulse and a probing free electron beam passing parallel to the surface at a distance of 50\,nm. We consider a metamaterial film consisting of $N=10$ vertical periods. (b) Calculated spatiotemporal dynamics of the graphene electron temperature assumed to be uniform across the thin metamaterial film. The spatial coordinate indicates the distance to the center of the axisymmetric optical Gaussian pulse (600\,nm beam width). (c,d) Spatiotemporal dynamics of the EELS signal for an electron frequency loss of 42\,THz, as obtained through full electromagnetic calculations (c) or using an effective medium model for the metamaterial (d). The occurrence of the topological transition of the isofrequency contour, taking place at $\Te \approx 3560\,$K for the optical frequency under consideration, is indicated by blue-dashed curves in (b-d), which enclose the spatiotemporal domain characterized by a hyperbolic response.}
\label{Fig3}
\end{figure}

In order to resolve the ultrafast spatial and temporal dynamics of the topological transition, an electron probe moving in free space, can be employed to spatially image the topological transition in the so-called aloof configuration \cite{paper149} after a short optical pulse pumping, as illustrated in Fig.\ \ref{Fig3}(a). This type of experiment can be performed with state-of-the-art ultrafast electron microscopes, relying on pulsed optical pumping and electron probing \cite{BFZ09,FES15,PLQ15,paper311,DNS19}. The probe electron can provide direct information about the topological transition through the electron energy-loss spectroscopy (EELS) signal, the probability of which is given by \cite{paper149}
\begin{align}
\frac{\Gamma(\omega)}{L}=\frac{2e^2}{\pi \hbar v^2}\int_{0}^{\infty}\frac{dk_y}{k_{\parallel}^2}{\rm Re} \left \{ \left(  \frac{k_y^2 v^2}{k_z^2 c^2}r_s-r_p \right) k_z {\rm e}^{2\ii k_z z_0} \right \},
\label{EELS}
\end{align}
where $L$ is the length of the electron trajectory, $v$ is the electron velocity, $z_0$ is the separation distance between the electron beam and the metamaterial surface, $k_{\parallel}=\sqrt{\omega^2/v^2+k_y^2}$, and $k_z=\sqrt{k_0^2-k_{\parallel}^2}$. We note that the loss probability given by Eq.\ (\ref{EELS}) bears a close relation to the momentum decomposition of LDOS along the electron trajectory \cite{paper102}. In what follows, we set the electron velocity to $v=0.5c$ ($\approx100\,$keV energy) and the separation to $z_0=50\,$nm. We further consider a Gaussian pump pulse of 100\,fs duration, 2\,mJ/cm$^2$ fluence, and 600\,nm beam width, which is realistic for lasers in the visible range. Following optical pumping of a metamaterial film composed of $N=10$ vertical unit cells, we calculate the resulting spatiotemporal dynamics of the electron temperature using a two-temperature model \cite{paper313,paper330}. The results are shown in Fig.\ \ref{Fig3}(b), where the in-plane radial distance is referred from the Gaussian beam center. The electron temperature is a maximum value $\sim 4000\,$K at the beam center immediately after pumping and then decreases as time evolves. This range of high electron temperatures is currently achievable in state-of-the-art experiments \cite{JUC13,GPM13,TSJ13}. In a way that is consistent with its intimate relation to the LDOS, the spatiotemporal dynamics of the EELS signal follows closely that of $\Te$ at a fixed frequency loss of 42\,THz, as shown in Fig.\ \ref{Fig3}(c,d), where the results obtained from the effective medium model (Fig.\ \ref{Fig3}(d)) match quite well those obtained from the full calculation (Fig.\ \ref{Fig3}(c)). The blue-dashed curves in Fig.\ \ref{Fig3}(b-d) enclose a spatiotemporal regime in which the isofrequency contour is transformed to be hyperbolic, giving rise to an enhanced EELS signal. Within $\sim1\,$ps timescale, the topology of the isofrequency contour transforms from elliptic to hyperbolic, and then back to elliptic, thus encompassing an ultrafast topological transition.

\begin{figure}
\noindent \begin{centering}
\includegraphics[width=0.5\textwidth]{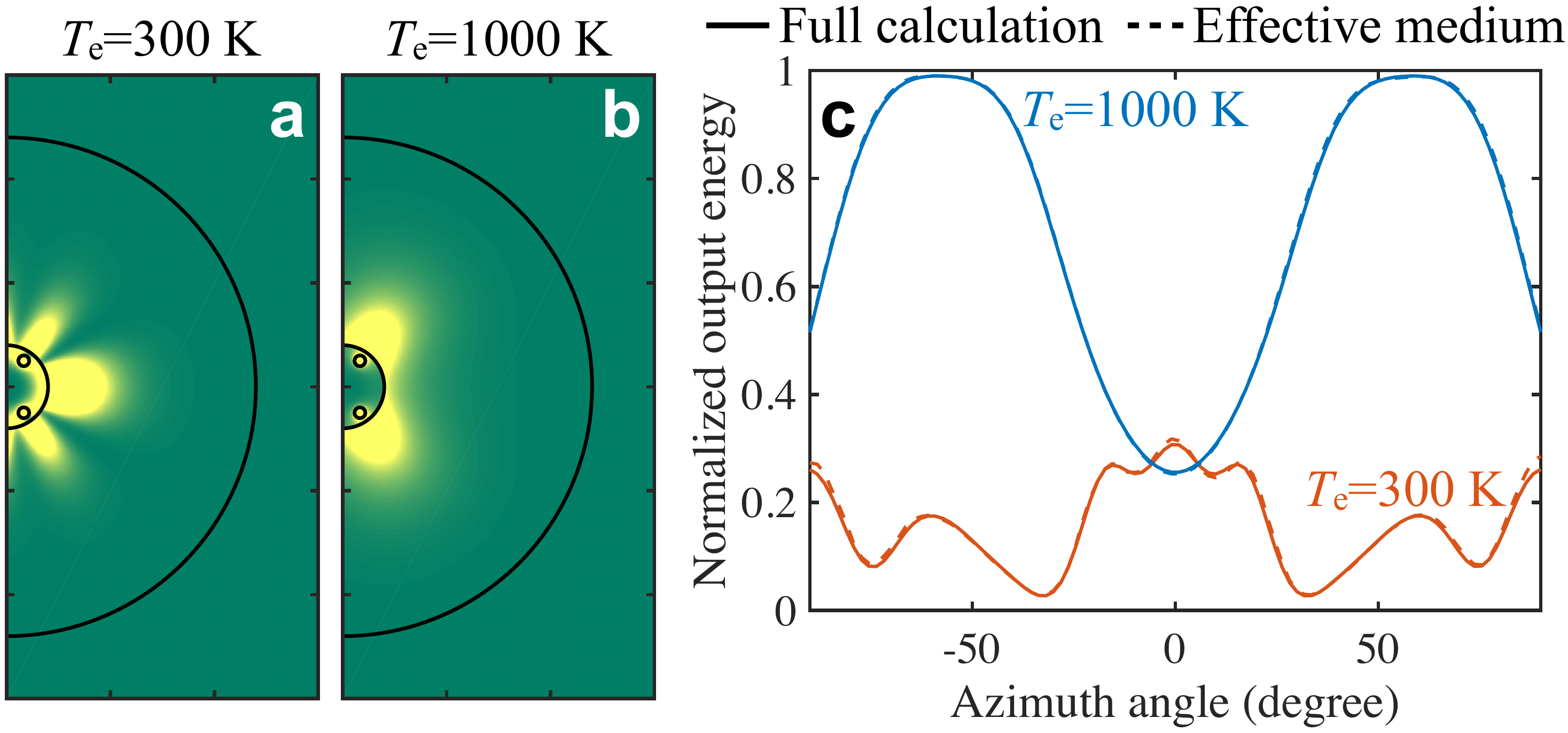}
\par\end{centering}
\caption{(a) Spatial distribution of the optical magnetic field in a plane perpendicular to a cylindrical metamaterial lens (bounded by the two black semicircles of radii $\lambda/5$ and $6\lambda/5$, respectively) at room temperature $\Te=300\,$K when it is excited by two magnetic current sources (represented by two small black circles close to the inner lens surface) separated by a subwavelength distance $\lambda/4$. We consider a light wavelength $\lambda=7.8\,\mu$m (i.e., 38.2\,THz frequency). The permittivity of the medium inside the inner circle and outside the outer one is taken to be 1 and 4, respectively. (b) Same as (a) when the graphene electron temperature in the metamaterial is raised to $\Te=1000\,$K. (c) Normalized far-field emitted energy as a function of azimuthal angle, spanning a range from $-90^\circ$ to $90^\circ$, at the two electron temperatures considered in (a,b). We compare results obtained from full electromagnetic simulations (solid curves) and the effective medium model (dashed curves).}
\label{Fig4}
\end{figure}

As a final example of application of the ultrafast topological phase transition discussed above, we investigate light steering and super-resolution imaging. It has been demonstrated that a perfect lens can be realized through negative refraction and amplification of evanescent waves in negative-index metamaterials \cite{P00}. Hyperbolic media with nearly flat isofrequency dispersion contours are also capable of producing subwavelength imaging of a point source \cite{JAN06,LLX07}. Here, we study the radiated power distribution emanating from two magnetic line current sources (black circles in Fig.\ \ref{Fig4}(a,b), separated by $\lambda/4$, where $\lambda=7.8\,\mu$m is the optical frequency corresponding to a 38.2\,THz frequency). These sources are placed in front of a cylindrical lens made of a curved version of the layered graphene/dielectric under consideration. At room temperature $\Te=300\,$K, the isofrequency contour is indeed hyperbolic at 38.2\,THz [see Fig.\ \ref{Fig1}(b)]. However, because of the curved nature of the isofrequency contour, several guided modes are excited with different wave vectors $k_{r}$ along the radial direction \cite{BS06}. As a result, multiple lobes show up in the spatial distribution of the magnetic fields [Fig.\ \ref{Fig4}(a)], which also transmit into the far-field [red curves in Fig.\ \ref{Fig4}(c)]. When $\Te$ is increased to 1000\,K, the so-called canalization condition \cite{BSI05,SE06_2} ${\rm Re} \{ \epsilon_\theta \}\approx0$ is satisfied leading to a nearly flat isofrequency contour. Here, $\epsilon_\theta$ is the effective permittivity of the cylindrical lens along the azimuthal direction. This results in two highly directive lobes in the spatial distribution of the magnetic field [Fig.\ \ref{Fig4}(b)], thus demonstrating that the cylindrical lens is capable of ultrafast beam steering driven by the transient elevation of $\Te$-rising upon optical pulse pumping. Additionally, two clear angular peaks associated with those two individual point sources can be identified in the far-field regime [see blue curves in Fig.\ \ref{Fig4}(c)], further supporting the potential for far-field subwavelength imaging in a dynamical and ultrafast manner. Once more, the azimuthal distribution of the far-field signal obtained from full numerical simulations (solid curves) matches well the result obtained from the effective medium model (dashed curves), as shown in Fig.\ \ref{Fig4}(c). Our results further suggest a novel subwavelength image encoding/decoding mechanism, whereby an elevated electron temperature induced by optical pumping is the key to resolve encoded subwavelength images in the far-field regime.

\section*{Conclusion}

In summary, we have shown that transient heating can significantly modify the topology of the isofrequency dispersion contours in metamaterials formed by layered graphene/dielectric stacks by exploiting the remarkably small electronic heat capacity of graphene, which allows us to efficiently elevate the electron temperature through ultrafast optical pumping. By examining the spatiotemporal dynamics of the EELS signal obtained by using an electron-beam probe, we have found that the contour topology can transform between elliptic and hyperbolic shapes within a sub-picosecond timescale, that is, the characteristic time over which the elevated electron temperature evolves in real space. We can thus manipulate the LDOS enhancement in the proximity of the metamaterial, reaching variations in the calculated LDOS of a few orders of magnitude at the frequency in which this topological transition appears. Additionally, this type of transition enables ultrafast beam steering, which we have illustrated by illuminating a cylindrical lens made of metamaterials with two line sources, in which dynamical far-field subwavelength imaging has been demonstrated. Our findings open a promising route toward ultrafast control of light emission, beam steering, and optical image processing.

\section*{Acknowledgements}
This work has been supported in part by the ERC (Advanced Grant 789104-eNANO), the Spanish MINECO (MAT2017-88492-R and SEV2015-0522), the Catalan CERCA Program, and Fundaci\'o Privada Cellex. R.A. acknowledges the support of the  Alexander von Humboldt Foundation through the Feodor Lynen Fellowship. R.W.B. acknowledges support through the Natural Sciences and Engineering Research Council of Canada, the Canada Research Chairs program, and the Canada First Research Excellence Fund.

\bibliographystyle{apsrev}

\end{document}